\documentclass{JHEP3}
\preprint{LU-TP 06-10\\
  hep-ph/0602138}
  
\usepackage{cite}
\usepackage{epsfig}
\usepackage{color}
\let\normalcolor\relax
\usepackage{graphics}
\usepackage{axodraw}
\usepackage{inputenc}
\usepackage{xspace}
\inputencoding{latin1}
\renewcommand\email[1]{{\scriptsize\tt\href{mailto:#1}{#1}}}

\newcommand{\ybar}{\ensuremath{\overline{y}}}
\newcommand{\xbar}{\ensuremath{\overline{x}}}
\newcommand{\kbar}{\ensuremath{\overline{k}}}
\newcommand{\mbar}{\ensuremath{\overline{m}}}
\newcommand{\lbar}{\ensuremath{\overline{l}}}

\newcommand{\eqref}[1]{eq.~(\ref{#1})\xspace}

\newcounter{enumct}
\renewenvironment{enumerate}{\begin{list}{\Roman{enumct}.}%
{\usecounter{enumct}\setlength{\topsep}{1mm}%
\setlength{\partopsep}{1mm}\setlength{\itemsep}{0mm}%
\setlength{\parsep}{1mm}}}{\end{list}}

\skip\footins = 1\bigskipamount plus 2pt minus 4pt                              

\title{\boldmath Gravitational Scattering in the ADD-model Revisited }

\author{Malin Sjödahl\\
  Dept.~of Theoretical Physics,
  S\"olvegatan 14A, S-223 62  Lund, Sweden\\
  E-mail: \email{Malin.Sjodahl@thep.lu.se}}
  
  \abstract{
The effects of the condition that the standard model particles live on a 
thin brane in the ADD-model is investigated. This is shown to result in
a cut-off independent Kaluza--Klein propagator for large distance scattering
and an odd number of extra dimensions. 
The matrix element corresponding to this propagator can also 
be Fourier transformed to position space, giving 
back the extra-dimensional version of Newton's law. 
For an even number of extra dimensions the corresponding propagator 
is found by requiring that Newton's law should be recovered.}

\begin{document}
 
\sloppy
 
\section{Introduction}
\label{sec:intro}

The ADD-model 
\cite{Arkani-Hamed:1998rs,Arkani-Hamed:1998nn,Antoniadis:1998ig} 
aims at explaining the 
hierarchy between the weak scale and the Planck scale. 
This is done by introducing extra, compactified,  dimensions 
in which only gravity is allowed to propagate. At distances small compared to 
the compactification radius, but large compared to the Planck length
and the thickness of the brane,
the gravitational force will be enhanced and
behave essentially as in a $4+n$ dimensional world,
where $n$ is the number of extra dimensions.
If the extra dimensions are large enough, the enhanced gravitational force 
opens up for the possibility of
gravitational scattering and black hole production at present, or soon 
upcoming, collider experiments.

To quantify the amount of gravitational interaction, the theory was put on a
perturbative field-theoretical basis in \cite{Han:1998sg,Giudice:1998ck}. 
The perturbations of the extra dimensional part of the metric enter 
as massive Kaluza--Klein (KK) modes in 
the Lagrangian. When these modes
are internal states they have to be summed over which lead the authors 
of \cite{Han:1998sg,Giudice:1998ck} to the following divergent propagator 
integral

\begin{equation}
  \label{eq:mf}
  \sum_{\lbar}{\frac{1}{-m_{\lbar}^2+k^2}} 
  \sim R^n \int{\frac{m^{n-1}}{-m^2+k^2}}dm.
\end{equation}
Here $\lbar$ enumerates the allowed momenta, $m_{\lbar}$, in the extra 
dimensions, $m$ is the absolute value of $m_{\lbar}$, $R$ is the 
compactification radius and $k^2$ is the 
momentum squared of the $3+1$-dimensional part of the propagator. 
(This object, without Lorentz structure, will somewhat sloppily 
be referred to as a propagator.)
For the above approximation to be valid, the compactification radius $R$
of the extra dimensions clearly have to be large compared to other 
relevant length scales.
As it stands, the integral \eqref{eq:mf} is explicitly divergent for $n\geq2$. 
However, when arriving at \eqref{eq:mf}, the physical condition that the 
standard model fields are confined to the brane was not
taken into account. 

Instead the problem of the divergent integral was approached in  
\cite{Han:1998sg,Giudice:1998ck}
by introducing a cut-off $M_s$, argued to be of the same order of magnitude 
as the fundamental Planck scale, $M_p$. (A physical motivation for a 
cut-off was later considered in \cite{Bando:1999di,Kugo:1999mf}
by the introduction of a brane tension, and various cut-off versions have 
been discussed in \cite{Giudice:2003tu}). 
For $n>2$, and exchanged momentum 
small compared to $M_s$, the Kaluza--Klein summation of t-channel 
(or s-channel) amplitudes then gave a propagator behaving as  

\begin{equation}
  \label{eq:pa}
  \frac{1}{n-2} R^n M_s^{n-2}
  \sim  \frac{1}{G_{N(4)}}\frac{1}{n-2}\frac{M_s^{n-2}}{M_p^{n+2}}
\end{equation}
where $G_{N(4)}$ is the ordinary 3+1-dimensional Newton's constant
related to $R$ and $M_p$ via $G_{N(4)}^{-1}\approx R^n M_p^{n+2}$.
(In \cite{Han:1998sg} $M_p$ and $M_s$ are taken to be the same in the 
calculation of the propagator.) In the Born approximation the cross 
section would then be given by
\cite{Atwood:1999qd}

\begin{equation}
  \label{eq:pa2}
  \frac{d\sigma}{dz} \sim 
  \frac{s^3}{(n-2)^2}\left(\frac{M_s^{n-2}}{M_p^{n+2}}\right)^2
  F(\mbox{spin},z),
\end{equation}
where $z$ is the cosine of the scattering angle in the center of mass system, 
$s$ the squared sum of the incoming particles momenta and $F$ 
a function taking spin dependence into account.

A different approach to calculating the influence of the Kaluza--Klein modes,
was presented in \cite{Arkani-Hamed:1998nn} when deriving Newton's law. 
In this case the summation of Kaluza--Klein modes was performed 
in the classical limit, \textit{after} Fourier transforming our 
normal momentum space to coordinate space 
\cite{Arkani-Hamed:1998nn}, giving rise to the expected $3+n$ dimensional 
version of Newton's law,

\begin{equation}
  \label{eq:Newton}
  \frac{V(r)}{m_1 m_2} 
  \sim 
  \int_{-\infty}^{\infty} d\mbar 
  \int_{-\infty}^{\infty} d\kbar 
  \frac{1}{\mbar^2+\kbar^2} e^{i \kbar \cdot  \xbar} 
  \sim
  \int_{0}^{\infty} dm m^{n-1} \frac{e^{-mr}}{r}
  \sim \frac{1}{r^{n+1}}
\end{equation}
where $r=|\xbar|$. 
Eq.~(\ref{eq:pa}) and \eqref{eq:Newton} may seem to contradict each other.
Especially, \eqref{eq:pa} gives the same form of the gravitational 
scattering regardless of the number of extra dimensions, namely a 
 $\delta$-function at  $r=0$. Equation 
(\ref{eq:Newton}), on the other hand, gives different scattering behavior 
for different number of extra dimensions.
We will see in the next section that $k$-dependent
correction terms to \eqref{eq:pa} are important for the classical limit
as the $1/r^{n+1}$-potential can not be recovered by Fourier 
transforming the non-relativistic matrix element corresponding to 
\eqref{eq:pa} to ordinary position space.  

As we like to implement the condition that standard model fields 
live on a brane via a Fourier transform, we should really add a
factor  $i \ybar \cdot  \mbar$, where $\ybar$ is the coordinate in the
extra dimensions, in the exponent and then integrate over a narrow 
distribution in $\ybar$.  However, 
as the $m$ integral is explicitly convergent when evaluated after the $k$ 
integral, the addition of $i \ybar \cdot \mbar$ in the exponent would 
not change the result in the classical limit when the distance is much 
larger than the brane thickness.
In essence, what will be done in this paper is nothing but calculating 
\eqref{eq:Newton} in the reversed order, starting with the $m$-integral. 

The Fourier transform to position space w.r.t. the extra dimensions 
is performed (for an odd number of extra dimensions) in section 
\ref{sec:KKsum}. In section \ref{sec:Fourier} we show that the 
large distance limit of the resulting propagator can be 
further Fourier transformed to position space w.r.t. our normal dimensions
giving back the extra-dimensional version of Newton's law. 
We also find the corresponding KK-''summed'' propagator for an 
even number of extra dimensions by requiring that Newton's law should be 
recovered. Finally we summarize and conclude in section \ref{sec:Conclusion}.

\section{Fourier transformation to position space in the extra dimensions}
\label{sec:KKsum}

As the standard model fields are assumed to live on the brane,
any measurement of gravitational scattering will be in position space w.r.t. 
the extra dimensional coordinates and in momentum space w.r.t. our ordinary
coordinates. We therefore search for the corresponding propagator.
To find it we introduce a coordinate $\ybar$, with absolute value $y$, 
in the extra dimensions. Later we will be interested in a narrow 
distribution around $\ybar=\bar{0}$ corresponding to a small extension of the 
standard model fields into the bulk.
Searching for a propagator which is in position space w.r.t. the 
extra dimensions also corresponds to imposing the condition of locality.
As the standard model fields live on a brane they will only be 
sensitive to KK modes which overlap with the brane. 

Fourier transforming \eqref{eq:mf} to 
position space with respect to the extra dimensions now gives 
for $n\geq3$ 

\begin{eqnarray}
  \label{eq:mf2}
  D(k,y)&=& \sum_{\lbar} 
  \frac{1}{-m_{\lbar}^2+k^2}e^{i2 \pi \lbar \cdot \ybar /R }\nonumber\\
  &\approx& \left(\frac{R}{2 \pi} \right)^n S_{n-1}
  \int_{0}^{\pi} \sin(\theta)^{n-2} d\theta \int_{0}^{\infty} dm
  \frac{m^{n-1}}{-m^2+k^2} e^{imy\cos(\theta)} \nonumber\\
  &=&\left(\frac{R}{2 \pi} \right)^n S_{n-1}
  \int_{-1}^{1} d \cos(\theta) (1-\cos(\theta)^2)^{(n-3)/2} 
  \int_{0}^{\infty} dm \frac{m^{n-1}e^{imy\cos(\theta)}}{-m^2+k^2}
\end{eqnarray}
where $S_{n-1}$ is the surface of a unit sphere in $n-1$ dimensions 
(from integration over the angles on which the integrand does not depend) 
and the factor $R^n/(2\pi)^n$ comes from the density of Kaluza--Klein modes.
For an odd number of extra dimensions this can, with $x=\cos(\theta)$, 
be rewritten as
\begin{eqnarray}
  \label{eq:mf3}
  D(k,y)&=&
  \left(\frac{R}{2 \pi} \right)^n S_{n-1}
  \int_{-1}^{1} d x (1-x^2)^{(n-3)/2} \int_{0}^{\infty} dm
  \nonumber\\
  & &\left(-\sum_{j=0}^{(n-3)/2} m^{n-3-2j}k^{2j} 
  +\frac{k^{n-1}}{-m^2+k^2} \right)e^{imyx}.
\end{eqnarray}
From this form we see that the terms in the sum are either $k$-independent,
and therefore correspond to $\delta$-functions at $r=0$ in ordinary position 
space, or are even powers of $k$ and correspond to derivatives of the 
$\delta$-function. (This is easily seen by Fourier transforming from 
$k$-space to $r$-space component by component.) 
These terms are therfore localized
to distance zero in ordinary space and can not be important in the 
classical region, where wave packages are well separated and do not overlap.

If we imagine the brane to have a finite narrow distribution (such as
a Gaussian) around 
$\ybar=\bar{0}$ in the extra dimensions, the $1/r^{n+1}$-law will
be modified at distances similar to the brane thickness.
We also note that the terms in the sum give (possibly) 
large but finite contributions.
These terms, which depend on the extension of the brane, are important 
when the wave functions overlap and may then give an interaction 
similar to that from \eqref{eq:pa2} (until the unitarity condition sets in) 
but they will not be further investigated here.
Instead we concentrate on the term relevant for large distance 
and large energy scattering (the classical limit)
coming from the last term in the integral. The contribution to the integral  
from this term, called $\hat{D}(k)$, is easily evaluated in the limit of 
small $y$ (corresponding to a narrow distribution) and is given by

 \begin{equation}
  \label{eq:Iky}
  \hat{D}(k)\approx \left(\frac{R}{2\pi} \right)^n \frac{\pi S_n}{2} 
  \left(\sqrt{-k^2}\right)^{n-2} i^{n+1}.
\end{equation}
It is easy to show 
that this result holds also for one extra dimension, and therefore for any 
odd number of extra dimensions.

The expression \eqref{eq:Iky}, which is in momentum space with respect to our 
normal dimensions, and in position space with respect to the extra 
dimensions, has the following properties: 

\begin{enumerate}
\item[i]{It gives back Newton's law, \eqref{eq:Newton}. 
This will be demonstrated in section \ref{sec:Fourier}}.

\item[ii]{It depends on the number of extra dimensions in a non-trivial 
way, such that, as the gravitational force increases faster with smaller 
distance in position space for many extra dimensions, this is reflected 
in a faster increase with larger $k$ in momentum space.}

\item[iii]{It does not depend on an arbitrary cut-off as long as the
cut-off ($\approx 1/($brane thickness$)$) is much larger than $k$. 
This implies that 
it is not dominated by metric perturbations of the scale $1/M_p$ 
for scatterings corresponding to much larger distances. This is the 
case for \eqref{eq:mf} integrated to $M_s \approx M_p$.}
  
\item[iiii]{It is the part of the propagator 
which is argued to contribute to the all order exponentiated eikonalized 
amplitude in \cite{Giudice:2001ce} (apart from what appears to be 
a sign misprint).}
\end{enumerate}

\section{Fourier transformation to position space in our ordinary dimensions}
\label{sec:Fourier}
To obtain the 3+1+n dimensional version of Newton's law we take the classical 
limit such that the energy is given by the mass, multiply with the 
coupling constant $4 \pi G_{N(4)}$ and Fourier transform 
\eqref{eq:Iky} to position space.  Using $\kappa=|\kbar|$ we have

\begin{eqnarray}
  \label{eq:Iky2}
  \frac{V(r)}{m_1m_2}&=& 
  4\pi G_{N(4)}\left(\frac{R}{2\pi} \right)^n \frac{\pi S_n}{2} i^{n+1}
  \frac{S_3}{2} \frac{1}{(2\pi)^3}\int_{-1}^{1} d\cos(\theta) 
  \int_{0}^{\infty} d\kappa 
  \kappa^2 \kappa^{n-2} e^{i\kappa r \cos(\theta) }\nonumber\\
  &=&4\pi G_{N(4)}\left(\frac{R}{2\pi} \right)^n \frac{\pi S_n S_3}{4(2\pi)^3} 
  i^{n+1}\frac{1}{ir}\int_{0}^{\infty} d\kappa 
  \kappa ^{n-1} (e^{i\kappa r}-e^{-i\kappa r}).
\end{eqnarray}
This can be evaluated by introduction of a small convergence factor
\begin{eqnarray}
  \label{eq:Ikyo}
  \frac{V(r)}{m_1m_2}&=&
  4\pi G_{N(4)} \left(\frac{R}{2\pi} \right)^n \frac{\pi S_n S_3}{4(2 \pi)^{3}}
  i^{n+1}\frac{1}{ir}
  \lim_{\epsilon \rightarrow 0}
  \left( \frac{d}{dr}\right)^{n-1}
  \int_{0}^{\infty} d\kappa \left[ 
    \frac{e^{i\kappa r-\epsilon \kappa}}{i^{n-1}}-
    \frac{e^{-i\kappa r-\epsilon \kappa }}{(-i)^{n-1}}
    \right] \nonumber\\
  &=&
  G_{N(4)} \left(\frac{R}{2\pi} \right)^n \frac{\pi S_n (4 \pi)^2}{4(2 \pi)^{3}} 
  i^{n+1}\frac{1}{ir}\frac{1}{i^{n-1}}
  \lim_{\epsilon \rightarrow 0}
  \left( \frac{d}{dr}\right)^{n-1}(-1)
  \left[
    \frac{-i}{r+i\epsilon}-\frac{i(-1)^{n-1}}{r-i\epsilon}
    \right]\nonumber\\
  &=&-G_{N(4)} \left(\frac{R}{2\pi} \right)^n S_n 
  \frac{\Gamma(n)}{r^{n+1}},
\end{eqnarray}
where the last step only is valid for an odd number of extra dimensions.
This is the same result as in \cite{Arkani-Hamed:1998nn} (apart from a
minus sign which is neglected in \cite{Arkani-Hamed:1998nn}), i.e. 
gravitational scattering enhanced by the large density of Kaluza--Klein 
modes, corresponding to a large coupling constant.

The strategy so far has been to start from the propagator and argue that we 
can get back Newton's law. Clearly this argument could be turned upside down. 
Using the result in \eqref{eq:Newton} \cite{Arkani-Hamed:1998nn}, 
we could alternatively search for the propagator giving the expected 
potential when Fourier transformed to position space. Again this would give 
us a propagator of the form $k^{n-2}$ for an odd number of extra dimensions.
For an even number of extra dimensions we settle with this argument, showing 
that the propagator

 \begin{equation}
  \label{eq:Ikye}
  \hat{D}(k) \approx \left(\frac{R}{2\pi} \right)^n \frac{S_n}{2} 
  (-1)^{\frac{n-2}{2}}  \left(\sqrt{-k^2}\right)^{n-2}\ln(-k^2)
\end{equation}
gives the desired form. The logarithm of a dimension-full quantity 
may seem disturbing. Replacing it by $\ln(-k^2/k_0^2)$ for some $k_0$ we
note that the $ k^{n-2}\ln(k_0^2)$-term would only contribute 
at $r=0$ when Fourier transformed to position space. 
This means that, just as in the case of an odd number of extra 
dimensions, we have a local contribution which depends on the brane. 
We also note that, just as for odd $n$, requiring
the standard model particles to live on a thin bane by introducing a 
(narrow) distribution in $y$-space we would have a (wide) distribution in
 $m$-space giving a (large) finite value for the propagator.

Fourier transforming the non-relativistic
version of \eqref{eq:Ikye} to position space we find the potential

\begin{eqnarray}
  \label{eq:Ikye2}
  \frac{V(r)}{m_1m_2}&=& 
  4\pi G_{N(4)}\left(\frac{R}{2\pi} \right)^n \frac{ S_n}{2}(-1)^{\frac{n-2}{2}}
  \frac{S_3}{2} \frac{1}{(2\pi)^3}\int_{-1}^{1} d\cos(\theta) 
  \int_{0}^{\infty} d\kappa 
  \kappa^2 \kappa^{n-2} \ln(\kappa^2) e^{i\kappa r \cos(\theta) }\nonumber\\
  &=& G_{N(4)}\left(\frac{R}{2\pi} \right)^n 
  \frac{S_n (4\pi)^2}{4(2\pi)^3}(-1)^{\frac{n-2}{2}} 
  \frac{1}{ir}\int_{0}^{\infty} d\kappa 
  \kappa ^{n-1} \ln(\kappa^2 )2 i \sin(\kappa r). 
\end{eqnarray}
For $n$ even we get after evaluating the integral \cite{Grad} 
 \begin{eqnarray}
  \label{eq:Ikye3}
  \frac{V(r)}{m_1m_2}
  &=& -G_{N(4)} \left(\frac{R}{2\pi} \right)^n S_n 
  \frac{\Gamma(n)}{r^{n+1}}.
\end{eqnarray}
This is the multidimensional version of Newton's law.
We also note that \eqref{eq:Ikye} is the part of the propagator argued to 
contribute to the all order eikonalized amplitude in \cite{Giudice:2001ce}.

 

\section{Conclusion and outlook}
\label{sec:Conclusion}
We have shown that 
requiring the standard model particles to live on a finite brane,
leads to a convergent result for the KK propagator.
The part of the propagator which is relevant for large energy and large 
distance scattering can also (at least for an odd number of extra dimensions) 
be Fourier transformed to position space w.r.t.\ our ordinary 
coordinates, giving back Newton's law. 
For an even number of extra dimensions we have found 
an expression for the KK-''summed'' propagator by requiring that we should get 
back Newton's law in the classical limit, when all coordinates are Fourier 
transformed to position space.

It should be pointed out that, although the tree-level amplitude is calculated
to a finite value, we can not use
\eqref{eq:Iky} and \eqref{eq:Ikye} for calculating large energy cross 
sections with the Born approximation. 
Instead an all order summation still has to be performed 
\cite{Giudice:2001ce}. 
Furthermore, the cross section obtained in \cite{Giudice:2001ce} is 
supported by the results here, as the finite results for the 
propagator here agrees with the parts argued to contribute to the 
eikonal cross section in \cite{Giudice:2001ce}.

\section*{Acknowledgments}
I am thankful to Leif Lönnblad, Johan Grönqvist, Gösta Gustafson, Pavel 
Kurasov and Johan Bijnens for useful discussions and comments on the 
manuscript and to Gian Giudice for making me aware of some of the cited 
articles.

\bibliographystyle{utcaps}  
\bibliography{references,refs}
\end{document}